\documentclass[12pt]{iopart}
\usepackage[dvips]{graphicx}
\usepackage{amssymb}
\usepackage{latexsym}

\begin{document}

\newcommand{\ii}{{\rm i}}
\newcommand{\dd}{{\rm d}}
\newcommand{\eps}{\epsilon}

\newcommand{\bC}{{\mathbb{C}}}
\newcommand{\bZ}{{\mathbb{Z}}}
\newcommand{\bR}{{\mathbb{R}}}
\newcommand{\bT}{{\mathbb{T}}}
\newcommand{\bN}{{\mathbb{N}}}

\newcommand{\braket}[2]{\langle#1|#2\rangle}
\newcommand{\bra}[1]{\langle#1|}
\newcommand{\ket}[1]{|#1\rangle}
\newcommand{\ave}[1]{\left\langle #1 \right\rangle}
\newcommand{\mod} {\mathop{\rm mod}}
\newcommand{\bfm}[1]{{\mbox{\boldmath$#1$}}}

\newcommand{\beq}{\begin{equation}}
\newcommand{\eeq}{\end{equation}}
\newcommand{\beqa}{\begin{eqnarray}}
\newcommand{\eeqa}{\end{eqnarray}}


\title{Egorov property in perturbed cat map}

\author{Martin Horvat$^{(1,2)}$, Mirko Degli Esposti$^{(2)}$}

\address{
  ${}^{(1)}$ Physics Department, Faculty of Mathematics and Physics,
  University of Ljubljana, Slovenia\\
  ${}^{(2)}$ Department of Mathematics, University of Bologna, Italy
}

\eads{\mailto{martin.horvat@fmf.uni-lj.si}, \mailto{desposti@dm.unibo.it}}
\begin{abstract}
We study the time evolution of the quantum-classical correspondence (QCC) for the well known model of quantised perturbed cat maps on the torus in the very specific regime of semi-classically small perturbations.
The quality of the QCC is measured by the overlap of classical phase-space density and corresponding Wigner function of the quantum system called quantum-classical fidelity (QCF).
In the analysed regime the QCF strongly deviates from the known general behaviour discussed in \cite{horvat:non:qcf06}, in particular it decays faster then exponential. Here we study and explain  the observed behavior of the QCF and the apparent violation of the QCC principle.
\end{abstract}
\submitto{\JPA}


\section{Introduction}
The quantum-classical correspondence (QCC) is the basic principle underlying  any physical quantisation of a classical system. According to this principle the quantum system should behave similar to the corresponding classical system with increasing energy or decreasing effective Planck constant.
The importance of QCC as a tool in the study of quantum systems  was recognised very early in the development of quantum mechanics with the Ehrenfest theorem and later by the introduction of semi-classical methods \cite{landauQM:book:91}. The study of QCC gave in the 80's birth to quantum chaology -- research area devoted to study the connections between dynamical properties of classical systems and corresponding quantum systems \cite{haake:book:01}.

The QCC can be explored and discussed using various tools and methods available in the theory of classical/quantum  systems. In a recent paper \cite{horvat:non:qcf06}, a phase space representation has been used to study the time evolution of QCC in generic chaotic systems on compact classical phase space.

The QCC is there quantified using the so called quantum-classical fidelity (QCF), namely the integrated overlap between the classical phase-space density and the corresponding Wigner function. It has been shown that in classically chaotic systems, after some initial plateau, the QCF decays exponentially in time with decay rate coinciding with the maximal Lyapunov exponent $\lambda$.
While it is common knowledge in the realm of quantum chaos that the phase space correspondence between classical and quantum mechanics drops down on the scale of Ehrenfest time $t_E\approx -\log\hbar/\lambda$, the exact dependence of initial plateau on dynamical properties and on Hilbert space dimension are still important  open questions that we aim to address here, at least in a special case. In particular, here we discuss QCC using QCF for the so called perturbed Arnold cat map\cite{arnold:book:89} on a torus $\bT^2 = [0,1]^2$.
The (unperturbed) cat map is a paradigmatic example of a classical uniformly hyperbolic chaotic systems. It has been one of the first extensively studied quantum maps \cite{hannay:physicaD:80} and since then it has been used several times to prove or disprove various conjectures concerning statistical  properties of  eigenfunctions for quantum system with strongly chaotic classical motion (see for example \cite{degli:book:03} and references therein). Because of the linearity of the classical motion, its quantum counterpart inherits a natural non generic number-theoretical  structure, reflected for example in the rigid distribution of eigenvalues and also in the so called {\it exactness} of the Egorov property, which roughly means that classical and quantum time evolution perfectly commute.
A generic behaviour of eigenvalues, i.e.  a good agreement with the predictions of random matrix theory, can be gained by perturbing the linear cat dynamics by composing the map for example with a time one flow generated by a given (global) Hamiltonian over the torus \cite{dematos:ann_phy:95, boesman:prslA:95}. \par
We are here interested in exploring the time evolution properties of these perturbed maps. In particular we aim to study how QCC decays in the presence of perturbation and especially the time scales of the initial plateau in QCF.
\section{Quantisation on a torus and Egorov property}
We recall here the basic facts of quantum mechanics over the torus
which we need in the paper, see \cite{degli:book:03} and \cite{dana:jpa:02} for further details. The system is quantised on a torus $\bT^2$ by introducing a position basis $\{\ket{q_n}:q_n=\frac{n}{N}\}_{n \in \bZ_N}$ and a momentum basis $\{\ket{p_m}:p_m=\frac{m}{N}\}_{m\in \bZ_N}$ in the Hilbert space ${\cal H}_N$ of dimension $N$. We apply periodic boundary conditions $\ket{q_{n+N}}=\ket{q_n}$ and $\ket{p_{m+N}}=\ket{p_m}$. The two basis are related by the discrete Fourier transform:
\beq
  \braket{q_n}{p_m} = \frac{1}{\sqrt{N}} e^{\ii \frac{2\pi}{N} nm}\>.
\eeq
Then according to Weyl-Wigner quantisation we associate an operator $\hat A$ to a classical observable $a$ defined over grid points ${\cal G}_N = \{x_{n,m}=(\frac{n}{2N},\frac{m}{2N})\}_{(n,m) \in \bZ_{2N}^2}$ on classical phase-space $\bT^2$ using following relations
\beq
  a_{n,m} = \tr\{\hat A  \hat A_{n,m}\}\>,\qquad
  \hat A = \hat Q_{\rm w} (a) = \sum_{(n,m)\in\bZ_{2N}^2} a_{n,m}\hat A_{n,m}\>,
\eeq
where $\hat A_{n,m}$ is called the point operator or the kernel of the Weyl-Wigner formalism
\beq
  \hat A_{n,m} = \frac{e^{\ii\frac{\pi}{N}nm}}{2\sqrt{N}}
  \sum_{k \in \bZ_N} e^{-\ii\frac{2\pi}{N}km} \ket{q_{n-k}}\bra{q_k}\>,
\eeq
We say that $a$ is phase space representation of operator $\hat A$
or $\hat A$ is quantisation of phase-space function $a$. The
phase-space representation of a density operator corresponding to a
pure state $\hat\rho=\ket{\psi}\bra{\psi}$ is the Wigner function
$W_\psi(n,m)$ defined as
\beq
  W_\psi(n,m) = \frac{e^{\frac{\ii\pi}{N}nm}}{2\sqrt{N}} \sum_{k \in \bZ_N}
  \braket{\psi}{q_{n-k}}\braket{q_k}{\psi} e^{-\ii \frac{2\pi}{N} k m}\>,
  \qquad
  (n,m) \in \bZ_{2N}^2\>.
\eeq
with normalisation $\sum_{(n,m) \in \bZ_{2N}^2} W_\psi(n,m)^2 = 1$.
Namely,
\beq
  \hat Q_{\rm w} (W_\psi) =\hat\rho=\ket{\psi}\bra{\psi}.
\eeq
Let us assume that $M:\mathbb{T}^2 \to \mathbb{T}^2$ is a classical
discrete, area preserving, map on the torus. Then it is possible to
associate to $M$ a corresponding quantum evolution operator $\hat
U:{\cal H}_N \to {\cal H}_N$. In the following, $N$ is always an even integer in order to avoid certain technicalities in quantisation (see \cite{degli:book:03} and references therein). The propagator $\hat U$ will satisfy an {\it Egorov estimate}, namely \cite{bievre:aippt:98}:
\beq
 \lim_{N\to\infty}
  \parallel
    \hat U^\dag \hat Q_{\rm w} (a) \hat U - \hat Q_{\rm w} (a \circ M)
  \parallel =  0\>.
  \label{eq:egorov_exact}
\eeq
In particular, if the classical map is a linear automorphisms i.e
$M({\bf x}) = {\bf M}.{\bf x}$, where matrix ${\bf M}\in SL(2,\bZ)$ (e.g. the cat map) then {\it Egorov is exact}:
\beq
  \hat U^\dag \hat Q_{\rm w} (a) \hat U =\hat Q_{\rm w} (a \circ M).
\eeq
\section {Cat map and breaking of Egorov property}
\begin{figure}[!htb]
  \centering
  \includegraphics[width=7.5cm]{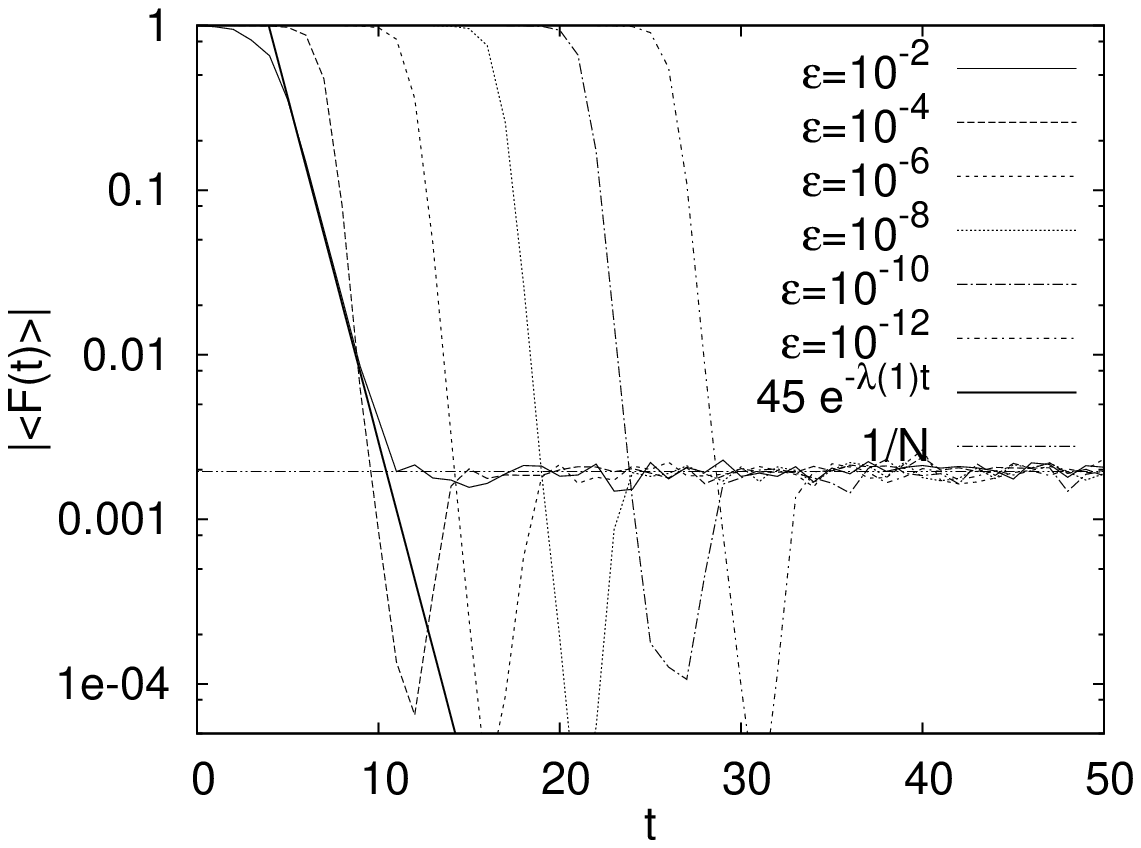}\hskip2pt%
  \includegraphics[width=7.5cm]{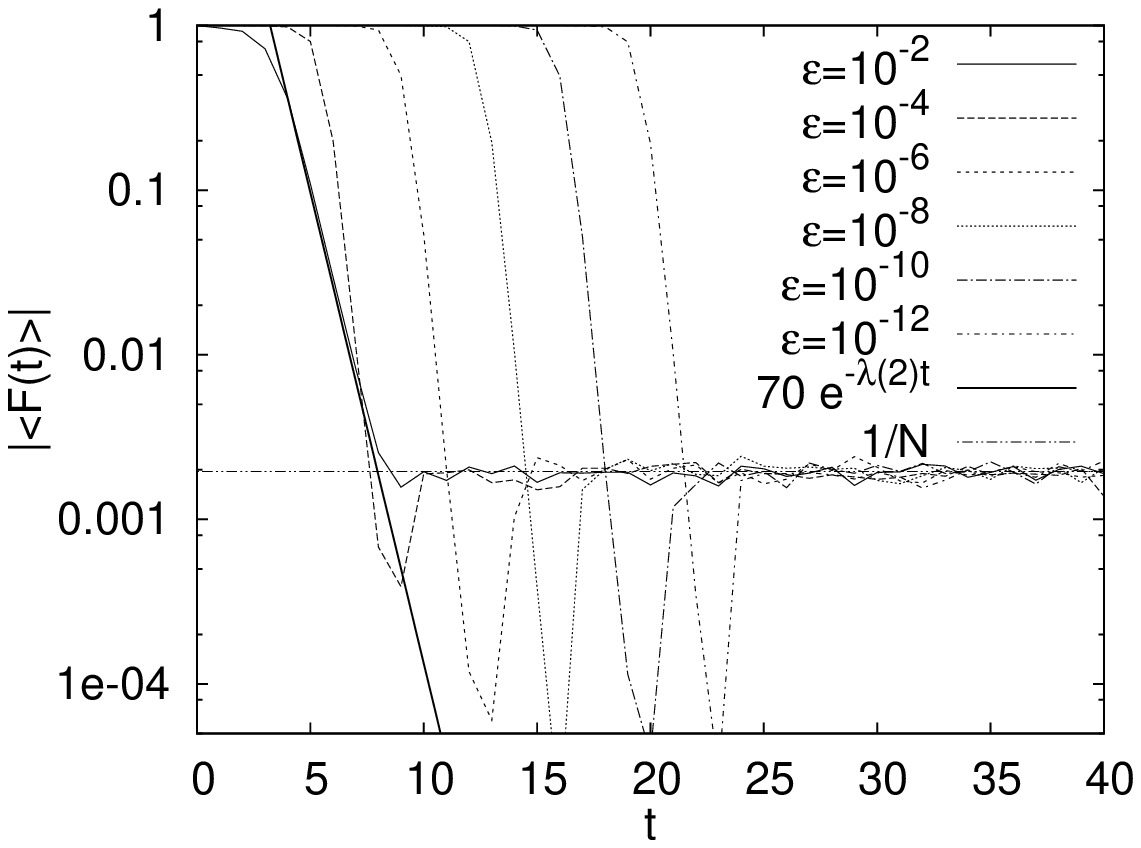}
  \hbox to15cm{\small\hfil(a)\hfil(b)\hfil}
\caption{The average QCF $F(t)$ with perturbation vector $\bfm{\eps}=(\eps,0,0)$ for different perturbations strength $\eps$ at Hilbert space dimension $N=512$ and $k=1,2$ (a,b). The average is taken over 100 initial Gaussian packets uniformly scattered over phase-space. }
\label{pic:qcf_F_ST}
\end{figure}

\begin{figure}[!htb]
  \centering
  \includegraphics[width=7.5cm]{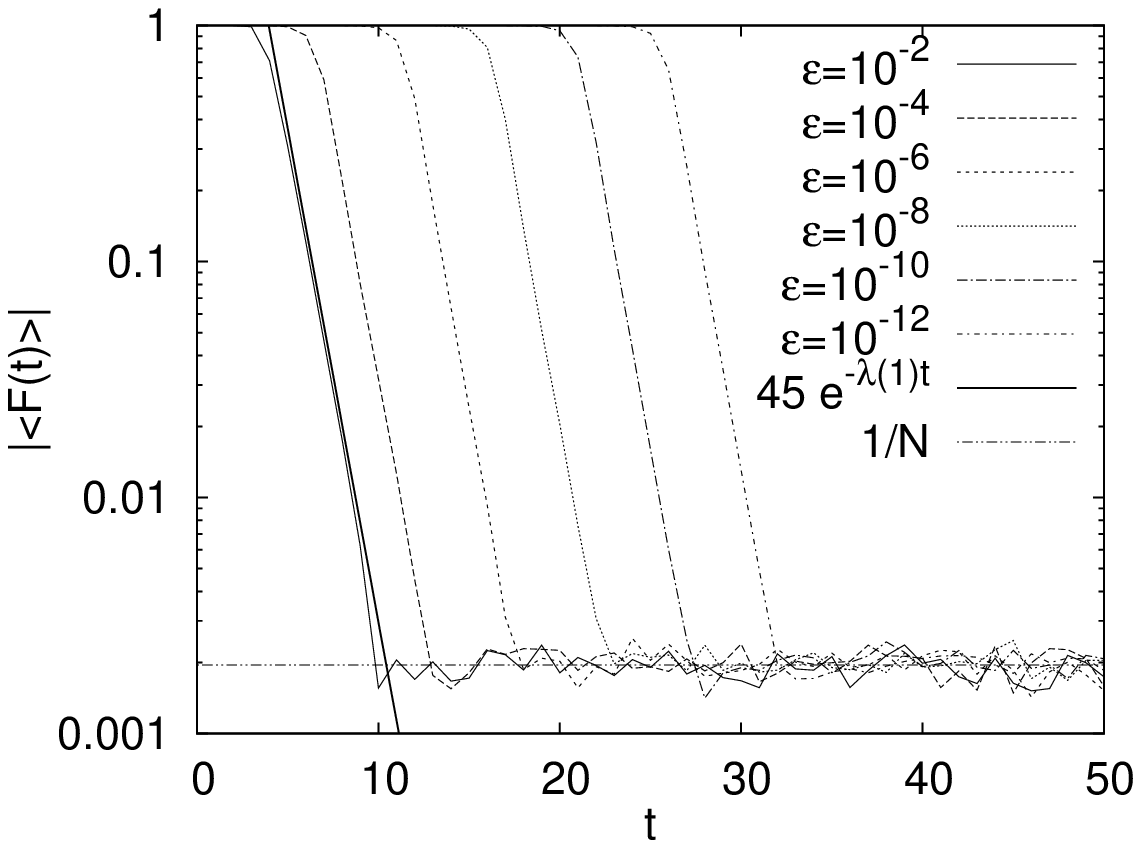}\hskip2pt%
  \includegraphics[width=7.5cm]{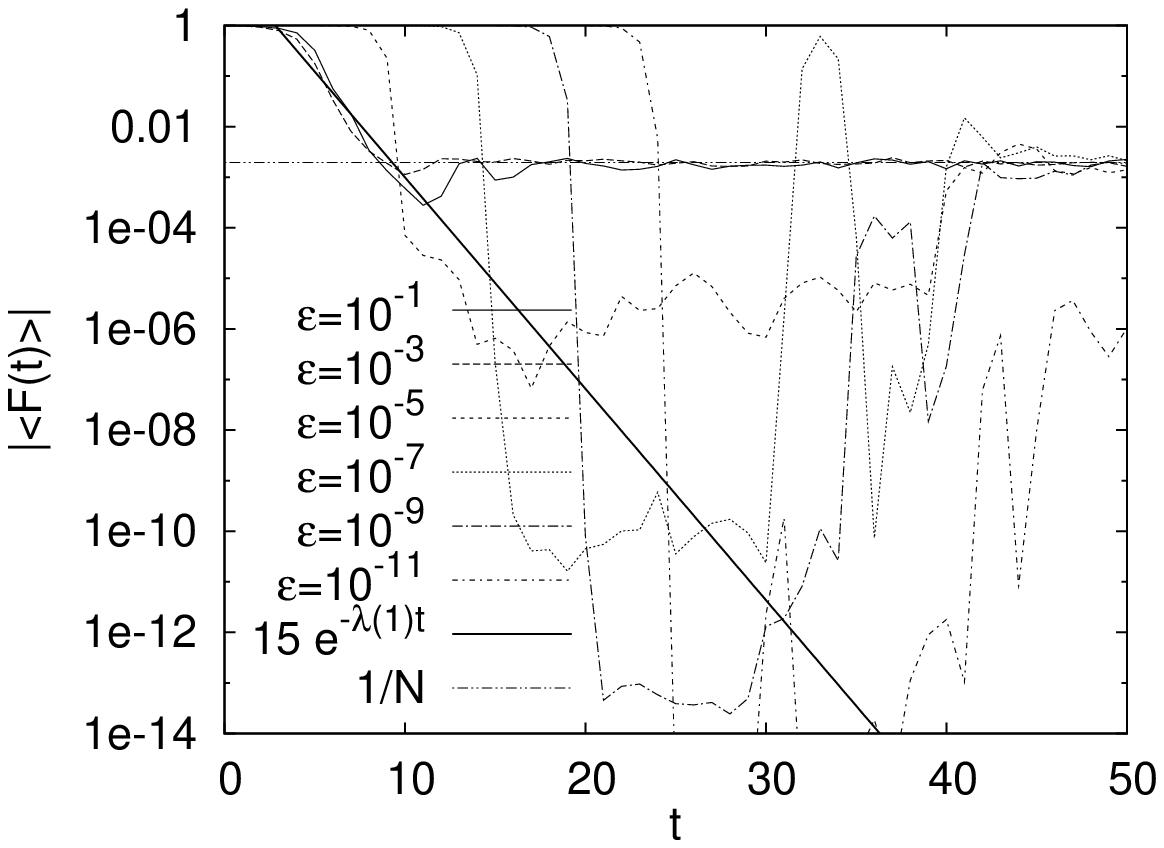}
  \hbox to15cm{\small\hfil(a)\hfil(b)\hfil}
\caption{The average QCF $F(t)$ for different perturbation strengths $\eps$ with perturbation vector $\bfm{\eps}=(0,\eps,0),~(0,0,\eps)$ (a,b) at Hilbert space dimension $N=512$ and $k=1$. For averaging see caption of fig. \ref{pic:qcf_F_ST}.}
\label{pic:qcf_F}
\end{figure}

%
%
The classical dynamics over the torus that we study here is given by the map $M:\bT^2\to\bT^2$
\beq
  (q,p)' = M(q,p)\>,
  \qquad
  \begin{array}{llll}
    p' &=& p + k q + \bfm{\eps}\cdot \dot{\bf V}(q) & \mod~1\cr
    q' &=& q + p'& \mod~1
  \end{array}\>,
\eeq
with $k\in\bN$, perturbation parameters $\bfm{\eps} = (\eps_0,\eps_1,\eps_2)\in\bR^3$ and perturbation function
\beq
  {\bf V} (q)
   = \left( \frac{1}{2} q^2, -\frac{1}{2\pi} \cos(2\pi q), q \right)\>.
\eeq
The quantum evolution operator corresponding to the perturbed cat map $\hat U$ can be written as
\beq
  \hat U
  = \exp \left(-\ii \frac{\pi}{N} \hat m^2\right)
    \exp
      \left(
      \ii k\frac{\pi}{N} \hat n^2 +
      \ii N \bfm{\eps}\cdot \hat {\bf V}
      \right)\>,
  \quad
  \hat{\bf V} = {\bf V} \left(\frac{2\pi}{N} \hat n\right)\>,
  \label{eq:evol_op}
\eeq
where we have for convenience introduced auxiliary operators
\beq
  \hat n \ket{q_n} = n \ket{q_n}\>,\qquad \hat m \ket{p_m} = m \ket{p_m}\>.
\eeq
If the perturbations are neglected $\bfm{\eps}=0$, we obtain the usual linear cat map system, where the map and evolution operator are denoted by
\beq
  M_{\rm c} = M|_{\bfm{\eps}=0}\>,\qquad
  \hat U_{\rm c} = \hat U|_{\bfm{\eps}=0}\>.
\eeq
As already remarked, it is well known that this system is Egorov
exact. Moreover, the classical cat-map is uniformly hyperbolic with the
Lyapunov exponent
\beq
  \lambda(k)
  = \log \left[\frac{1}{2}\left(k + 2  + \sqrt{k(k+4)} \right)\right]\>.
\eeq
For the coming analysis, it is convenient to write the classical map and the quantum evolution operators of the perturbed cat map as
\beq
  \hat U
  = \hat U_{\rm c} \exp\left(\ii N \bfm{\eps} \cdot \hat {\bf V} \right)\>,
  \qquad
  M
  = M_{\rm c} +  (\bfm{\eps} \cdot\dot{\bf V}, \bfm{\eps} \cdot\dot{\bf V})\>.
\eeq
We compare the classical and quantum evolution of the perturbed cat map in the classical phase space at some fixed dimension $N$ and perturbation $\bfm{\eps}$. The $\eps=\|\bfm{\eps}\|$ is refered to as perturbation strength. Here we are mainly interested in the particular case of semiclassical small perturbations $N \eps \ll 1$. More precisely, the classical system starts from a smooth probability distribution $\rho:\bT^2\to\bR$ resembling a Gaussian packet on phase space at the point $(q_0,p_0)$,
\beq
  \rho_{(q_0,p_0)} (q,p) = D_{N,(q_0,p_0)}
  \left (\sum_{\nu \in \bZ}  e^{-2\pi N (q-q_0 + \nu)^2}\right)
  \left(\sum_{\nu \in \bZ}  e^{-2\pi N (p-p_0 + \nu)^2}\right)\>.
  \label{eq:gauss_cl}
\eeq
where the scalar factor $D_N$ is pinned down by the normalisation
\beq
 \sum_{(n,m)\in \bZ_{2N}^2} \rho_{(q_0,p_0)}^2 (x_{n,m}) = 1\>,
\eeq
which has a simple leading term in the asymptotic approximation,  $N\to\infty$, reading
\beq
 \fl \hspace{5mm} D_{N,(q_0,p_0)} \asymp
  \frac{1}{\sqrt{N}}
    \left[
      \left(1 + 2\cos(4\pi N q_0) e^{-\pi N} \right)
      \left(1 + 2\cos(4\pi N p_0) e^{-\pi N} \right)
    \right]^{-\frac{1}{2}}\>.
\eeq
The quantum counterpart is initially in a coherent state $\ket{\phi}$ with a Wigner function $W_\psi$ similar to the classical distribution (see \cite{horvat:non:qcf06,nonnenmacher:06}):
\beq
  W_{\phi}(n,m) = \rho(x_{n,m}) + e^{-|O(N)|}\>,\qquad
  (n,m) \in \bZ^2_{2N}\>.
\eeq
We then let these two systems evolve up to time $t\in \bZ^*$ using equations
\beq
  \rho^t= \rho \circ M^{-t}\>,\qquad
  \ket{\phi^t} = \hat U^t \ket{\phi}\>,\qquad
  \hat\rho^t = \hat U^t \hat\rho \hat U^{-t}\>,
\eeq
and observe QCC between these two systems by calculating the overlap of the density $\rho^t$ and corresponding Wigner function $W_{\phi^t}$. The overlap is called quantum-classical fidelity (QCF) defined as
\beq
  F(t) = \sum_{(n,m) \in \bZ_{2N}^2} W_{\psi^t}(n,m) \rho^t(x_{n,m})
       = \tr\left\{\hat\rho^t\hat Q(\rho^t)\right\}\>  \le 1 + e^{-|O(N)|}
  \label{eq:qcf}
\eeq
Because the perturbed system is not Egorov exact, the QCF decreases with time. In figure \ref{pic:qcf_F_ST} we show numerically obtained the decay of average QCF $\ave{F(t)}$ for different $k$ and perturbation strengths $\eps$ using perturbation vector $\bfm{\eps}=(\eps,0,0)$, where $\ave{\bullet}$ denotes the uniform average taken over the initial positions of the  coherent packet.
The QCF  does not decay up to some time called the breaking time $t_{\rm br}$, which increases with decreasing perturbation. Beyond $t_{\rm br}$ the QCF decays ''very fast'' (as we will argue, faster then exponential) and eventually converges to the ergodic plateau given by $1/N$: the decay is in fact  visually faster than the generally expected exponential Lyapunov decay $\ave{F(t)} \sim \exp(-\lambda t)$ \cite{horvat:non:qcf06}, which is inserted in the figures. Basically the same scenario happens in other choices of perturbation vectors $\bfm{\eps}$ as we can see in figure \ref{pic:qcf_F}.
In the case of  constant classical perturbation $\bfm{\eps}=(0,0,\eps)$  shown in figure \ref{pic:qcf_F}.b, the correspondence is broken mainly by a rigid shifting of the deformed packets in the quantum and classical picture. Therefore the convergence towards the ergodic plateau is less smooth as in other cases. At this this point it is difficult to deduce the correct functional form  of the QCF $F(t)$.  Nevertheless, in the following we present a theoretical explanation of these numerical observations, disclosing  the super-exponential  nature of the QCF decay in this particular regime of perturbation.\par
\begin{figure}[!htb]
  \centering
  \includegraphics[width=7.5cm]{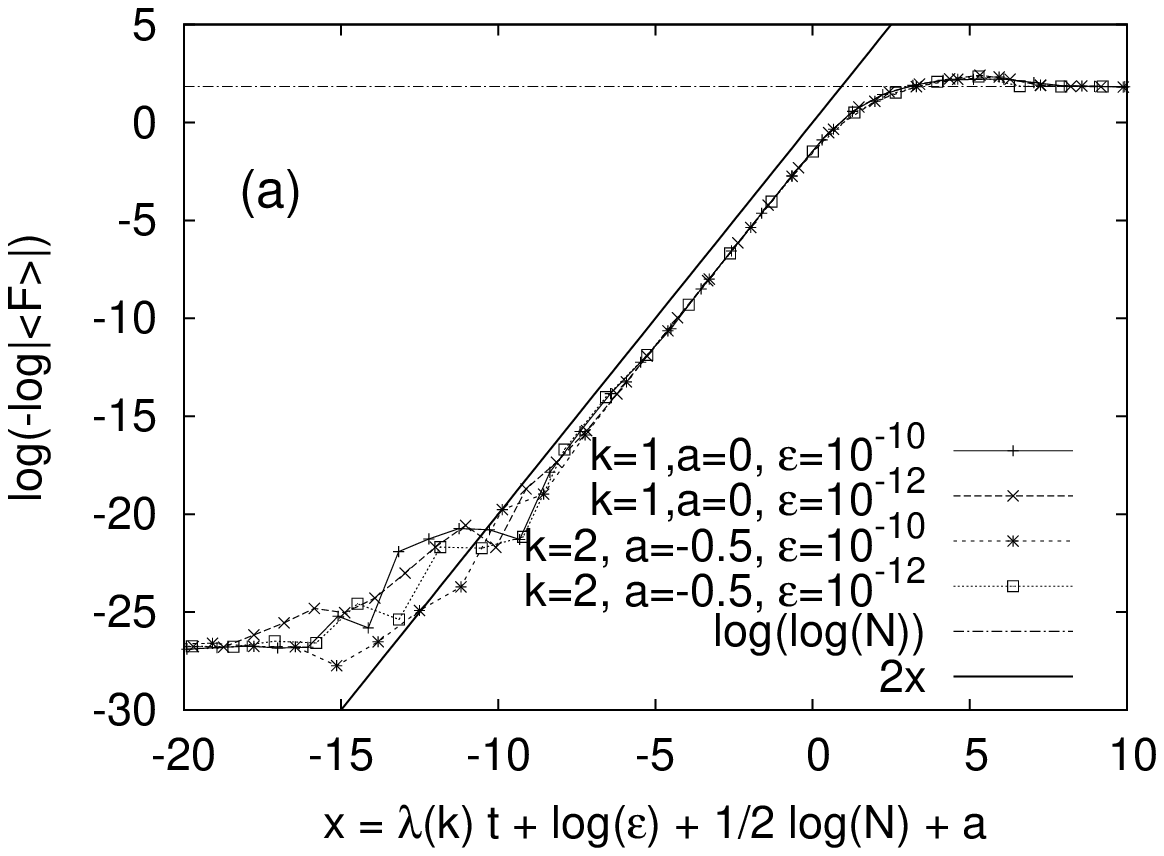}%
  \includegraphics[width=7.5cm]{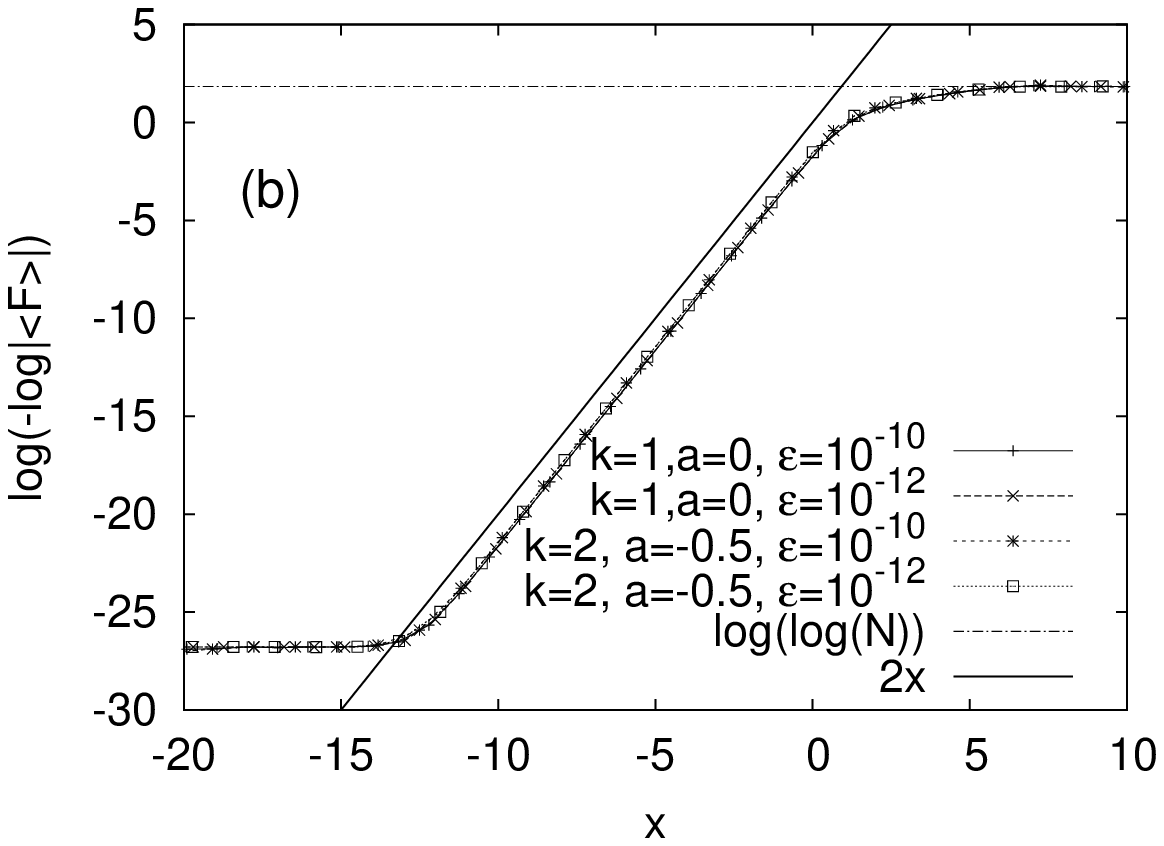}
  \includegraphics[width=7.5cm]{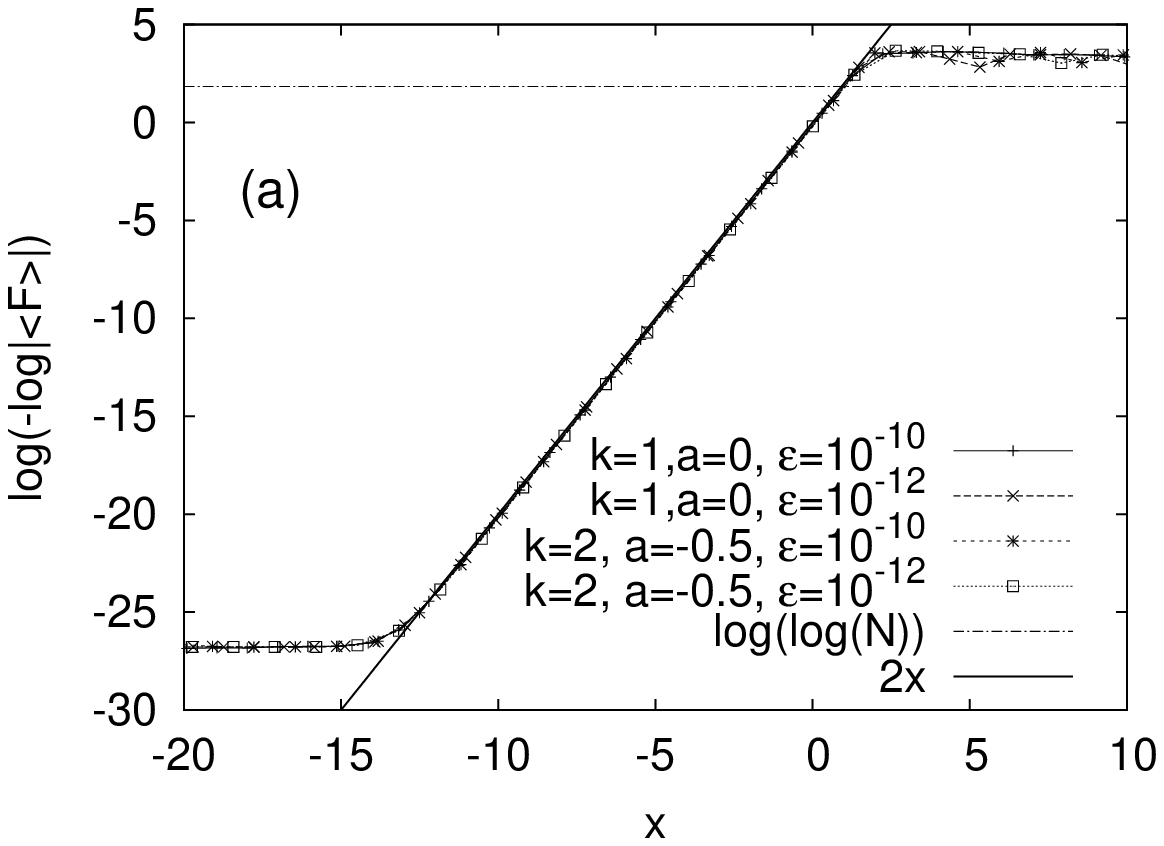}%
  \caption{The representation of average QCF $\ave{F(t)}$ evolution calculated using perturbation vector $\bfm{\eps}$ equal to $(\eps,0,0)$ (a), $(0,\eps,0)$ (b) and $(0,0,\eps)$ (c) for different $\eps$ and $k=1,2$ at the dimension $N=512$, where data presented in figs. \ref{pic:qcf_F_ST} and \ref{pic:qcf_F} is also considered. For averaging see caption of fig. \ref{pic:qcf_F_ST}. }
  \label{pic:qcf_cmp}
\end{figure}

We are interested in the evolution of the QCF in the limit of small perturbations $\eps\to 0$. In this regime we examine the time $t_{\rm br}(p)$ on which average QCF $\ave{F(t)}$ drops below some value $p$:
\beq
  t_{\rm br} (p) = \min\{t \in \bZ^*: \ave{F(t)} < p\}\>,
  \label{eq:time_break}
\eeq
where the average $\ave{\bullet}$ is taken uniformly over positions of the initial Gaussian packets. It is meaningful to express the dynamics relative to the unperturbed cat map writing
\beq
  \rho^t = \rho_{\rm c}^t + \delta \rho^t \>,\qquad
  \hat\rho^t = \hat\rho_{\rm c}^t + \delta\hat\rho^t\>,
  \label{eq:state_pert}
\eeq
where the dynamics of the cat map case is given by
\beq
  \hat\rho_{\rm c}^t = \hat U^t_{\rm c}\hat\rho\hat U^{-t}_{\rm c}\>,
  \qquad
  \rho_{\rm c}^t = \rho \circ M_{\rm c}^{-t}\>.
\eeq
and due to Egorov property these are connected by
\beq
  \hat\rho_{\rm c}^t = \hat Q(\rho_{\rm c}^t)\>.
\eeq
By inserting ans\"atze (\ref{eq:state_pert}) into the formula (\ref{eq:qcf}) we obtain QCF expressed in terms of deviations from the unperturbed case:
\beq
  F(t) = 1
  + \tr\left \{ \delta \hat\rho^t \hat Q(\rho_{\rm c}^t)  \right\}
  + \tr\left \{ \hat \rho_{\rm c}^t \hat Q(\delta \rho^t) \right\}
  + \tr\left \{ \delta \hat\rho^t  \hat Q(\delta \rho^t)  \right\}\>.
  \label{eq:qcf_dev}
\eeq
Due to existence of the Egorov property in the cat map, the approximated QCF can be expressed in terms of the quantum fidelity $F_{\rm q} (t)$ \cite{gorin:phy_rep:06} and the classical fidelity $F_{\rm c} (t)$ \cite{veble:prl:04} as
\beq
  F(t) = |F_{\rm q}(t)|^2  +  F_{\rm c} (t) - 1
           + \tr\left \{ \delta \hat\rho^t  \hat Q(\delta \rho^t)  \right\} \>,
  \label{eq:qcf_qf_cf}
\eeq
where $F_{\rm q}$ and $F_{\rm c}$ are here written as
\beq
  F_{\rm q}(t) = \bra{\phi} \hat U^{-t} \hat U_{\rm c}^t \ket{\phi}\>,\quad
  F_{\rm c}(t) = \sum_{(n,m) \in \bZ_{2N}^2}
  \rho(M^{-t}(x_{n,m})) \rho(M_{\rm c}^{-t}(x_{n,m}))\>.
\eeq
The relation (\ref{eq:qcf_qf_cf}) is very instructive and helps to understand the behaviour around the initial plateau, but it seems to us that the study of the plateau itself was greatly avoided in the past. In the following we discuss the second term and the third term in (\ref{eq:qcf_dev}) denoted by
\beqa
  I_1
  &=& \tr\left \{ \hat \rho_{\rm c}^t \hat Q(\delta \rho^t) \right\}
  = \sum_{(n,m) \in \bZ_{2N}^2}
    \rho_{\rm c}^t(x_{n,m}) \delta \rho^t(x_{n,m}) \>,
  \label{eq:term1}\\
  I_2
  &=& \tr\left \{ \delta \hat\rho^t \hat Q(\rho_{\rm c}^t)  \right\}
  = \tr\left \{ \delta \hat\rho^t \hat \rho_{\rm c}^t \right\}\>.
  \label{eq:term2}
\eeqa
The last term in (\ref{eq:qcf_dev}) and (\ref{eq:qcf_qf_cf}) are the second order corrections, which we do not discuss in detail.  In order to understand $I_1$ (\ref{eq:term1}) we discuss the deviation between trajectories of a chaotic and ergodic map $\phi = M^{-1}:\bT^2\to\bT^2$ and of its perturbation $\phi+\delta \phi = (M+\delta M)^{-1}$, starting at the same point $x$. The deviation is defined as
\beq
  \delta \phi_t (x) := (\phi + \delta \phi)^t(x) - \phi^t(x)\>,\qquad
  \phi^{t+1} (x) = \phi^t (\phi(x))  \>.
\eeq
and obeys in the limit $\delta \phi \to 0$ the following recursion
\beqa
  \delta \phi_{t+1}(x)
  &=& (\phi + \delta \phi)(\phi^t(x) + \delta \phi_t(x)) - \phi^{t+1}(x)\>, \\
  &\doteq& (\nabla \phi)(\phi^t(x)) \delta \phi_t(x) + \delta \phi(\phi^t(x))\>,
\eeqa
where we have neglected second order corrections. By iterating this equations from a given initial position $x$, the deviation is written as a series
\beq
  \delta \phi_t(x)
  = \sum_{k=0}^{t-1}
  \left[\prod_{l=k}^{t-1}(\nabla \phi)(\phi^l(x))\right]
  \delta \phi (\phi^{k-1}(x)) + \delta \phi(\phi^{t-1}(x))\>.
\eeq
Then by taking into account that  map is chaotic and ergodic with Lyapunov exponent $\lambda$, we get in the limits $t\to \infty$ and $\delta \phi \to 0$, applied in given order, the leading contribution of the deviation expressed as
\beq
   \delta \phi_t(x)
    = O(\delta \phi)\, e^{\lambda t}\>,\qquad
   \ave{\|\delta \phi_t(x)\|}\approx \eps A e^{\lambda t}\>.
\eeq
where $\ave{\bullet}$ denotes the uniform average over initial positions $x$. The constant $A \in \bR$ depends only on the type of perturbation and dynamical properties of the map. By plugging this result into expression $I_1$ (\ref{eq:term1}) we obtain
\beq
  I_1
  = \sum_{(n,m) \in \bZ_{2N}^2}
    \rho (x_{n,m})
     \rho \left(x_{n,m} + \delta \tilde\phi_t(x_{n,m})\right) -1\>,
  \quad
  \delta \tilde  \phi_t = \delta \phi_t \circ M^t\>,
\eeq
where we have used that the cat map $M$ conserves the grid $G_N$: $M(G_N) = G_N$. By taking into account the explicit form of $\rho$ (\ref{eq:gauss_cl}) and  considering only the behaviour about the central point of the Gaussian packet the above expression is approximated as
\beq
 \fl\hspace{1cm} I_1
  \approx
  \exp\left(-\pi N \|\delta \tilde \phi_t(q_0,p_0)\|^2\right) - 1\>,\quad
  \ave{I_1} \approx \exp (-\pi N A^2 \eps^2 \exp (2\lambda t))\>,
  \label{eq:term1_final}
\eeq
with $\ave{\bullet}$ representing the uniform average over position of the initial coherent packet. The approximation is meaningful up to times $\eps N^{\frac{1}{2}}\exp(\lambda t) = O(1)$, when deformation of the packets can be neglected. This is especially appropriate to described the case of constant classical perturbation. In the limit of small perturbations the leading term in expression $I_1$ scales with time and perturbation as $O(N\eps^2\exp(2\lambda t))$, where the changes of QCF are small. The behaviour of the expression $I_1$ is obtained by considering the fact that
\beq
  \hat U^t = \hat U_{\rm c}^t + \ii N \bfm{\eps} \cdot \sum _{k=1}^t
      \hat U_{\rm c} ^k \hat {\bf V} \hat U_{\rm c}^{t-k} +
      O\left((N \|\bfm{\eps}\hat {\bf V}\|)^2 t\right)\>,
\eeq
which yields
\beqa
  \delta\hat \rho^t
  &=&
  \ii N \bfm{\eps}\cdot \left[ \hat {\bf S}^t \hat \rho_{\rm c}^t -
                           \hat \rho_{\rm c}^{-t} \hat{\bf S}^{-t}\right]
  +
  O\left((N \|\bfm{\eps}\hat {\bf V}\|)^2 t\right)\>,\\
  \hat {\bf S}^t
  &=& \sum_{k=1}^t \hat U_{\rm c}^{-k} \hat {\bf V} \hat U_{\rm c}^k\>.
  \label{eq:quant_dev}
\eeqa
By plugging this into $I_2$ (\ref{eq:term2}) we get
\beq
  I_2
  = 2 N \bfm{\eps}\cdot
  \Im \left\{ \tr \left\{ \hat {\bf S}^t \hat\rho_{\rm c}^t\right\} \right\}
  + O\left((N \|\bfm{\eps}\hat {\bf V}\|)^2 t\right)\>.
  \label{eq:term2_final}
\eeq
By assuming that $\lim_{t\to\infty}t^{-1}\hat {\bf S}^t \neq 0$ we see that the leading term in $I_2$ scales as $O(N\eps t)$ in time. Then by considering results $I_1$ (\ref{eq:term1_final}) and $I_2$ (\ref{eq:term2_final}) we get the leading order contributions to QCF reading
\beq
  F(t) \approx
  2 N \bfm{\eps}\cdot
  \Im \left\{ \tr \left\{ \hat {\bf S}^t \hat\rho_{\rm c}^t\right\} \right\}
  +
  \exp\left(-\pi N \|\delta \tilde \phi_t(q_0,p_0)\|^2\right)\>.
  \label{eq:qcf_res}
\eeq
In the limit of the small perturbations the last term in (\ref{eq:qcf_res}) is the dominant. This is supported also numerically as we demonstrate in the figure \ref{pic:qcf_cmp}, where we show $G(t) = \log(-\log(\ave{F(t)})$ for different perturbation vectors $\bfm{\eps}$, perturbation strength $\eps$ and $k$.
We see that the average QCF evolves following the curve $G(t) \approx 2\lambda(k)t + {\rm const}.$ according to the dominant term in QCF  $I_1(t)$ (\ref{eq:term1_final}) almost up to the time, when QCF intersects the ergodic plateau given by $G(t)\approx \log\log N$. The plot $G(t)$ has an initial plateau due to finite arithmetic. We conclude that the QCF decays in average towards the ergodic plateau faster than exponentially as
\beq
  \ave{F(t)} = \exp(-|O(\exp(|O(t)|))|)\>.
\eeq
The expression for QCF (\ref{eq:qcf_res}) obtains in the limit $\sqrt{N} \eps \exp(\lambda t) \ll 1$ a simple scaling form
\beq
  \ave{F(t)} = 1 - O(N \eps^2 e^{2\lambda t}) + O (N \eps t)\>.
  \label{eq:qcf_scaling}
\eeq
where the first non-constant term is dominant in $F(t)$. In this perturbation approach we can approximate $t_{\rm br}(p)$ (\ref{eq:time_break}) for fixed $1-p\ll 1$ as
\beq
  t \approx \frac{\log(1- p) - \log (N\eps^2)}{2\lambda}\>.
  \label{eq:trb_gen}
\eeq
which in the limit of infinitesimal perturbations obtains following asymptotic form
\beq
  \lambda t_{\rm br} \asymp -\log \eps \>,\qquad \eps \to 0 \>.
  \label{eq:tbr_eps}
\eeq
We see that at fixed $N$ and $p$ the time depends only on Lyapunov exponent $\lambda$ and perturbation strength $\eps$.
\section {Numerical result on the breaking time}
In the following we present numerical results of the breaking time $t_{\rm br}$ in our perturbed cat map. We explore in particular its dependence on the perturbation strength $\eps$ and the Hilbert space dimension $N$.\par
\begin{figure}[!htb]
  \centering
  \includegraphics[width=7.5cm]{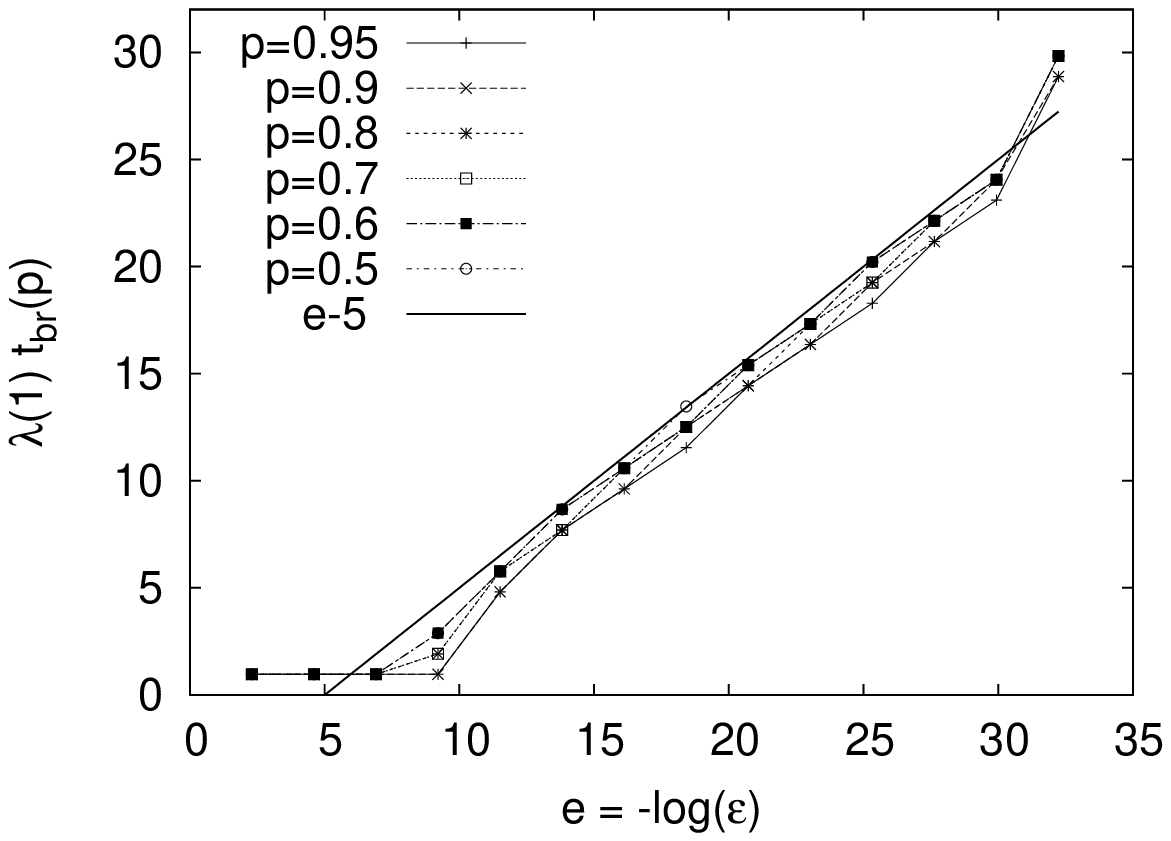}
  \includegraphics[width=7.5cm]{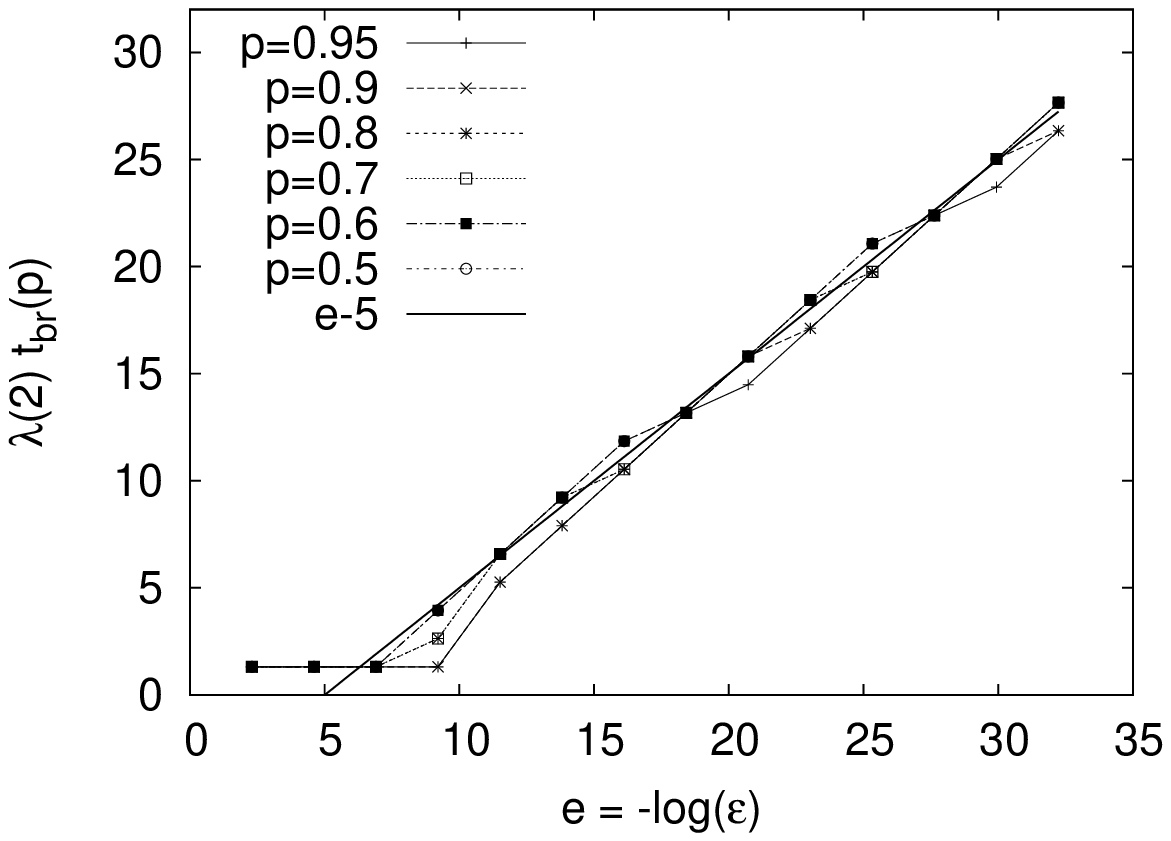}
  \hbox to15cm{\hfil(a)\hfil(b)\hfil}
\caption{The dependence of $t_{\rm br}$ on perturbation strength $\eps$ in the cases $k=1$ (a) and $k=2$ (b) by using perturbation vector $\bfm{\eps}=(\eps,0,0)$ at $N=512$ .}
\label{pic:qcf_T_ST}
\end{figure}
\begin{figure}[!htb]
  \centering
  \includegraphics[width=7.5cm]{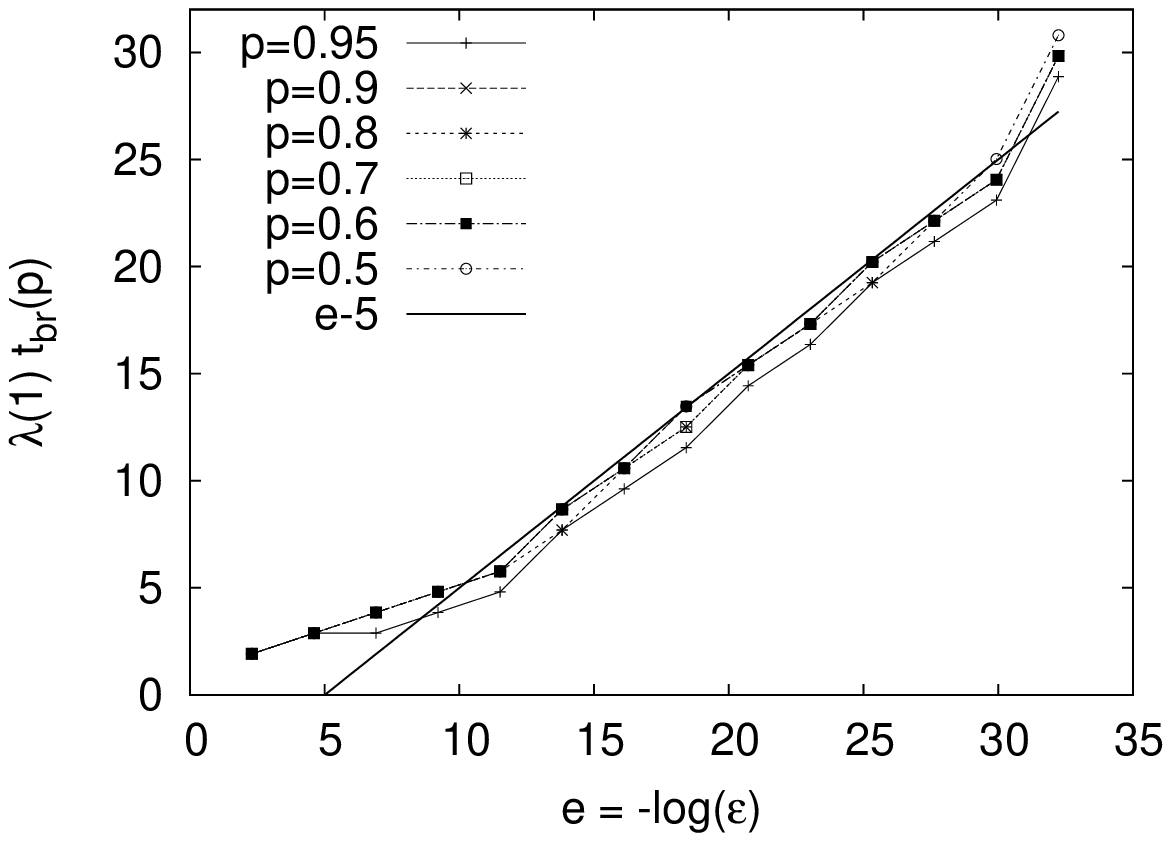}%
  \includegraphics[width=7.5cm]{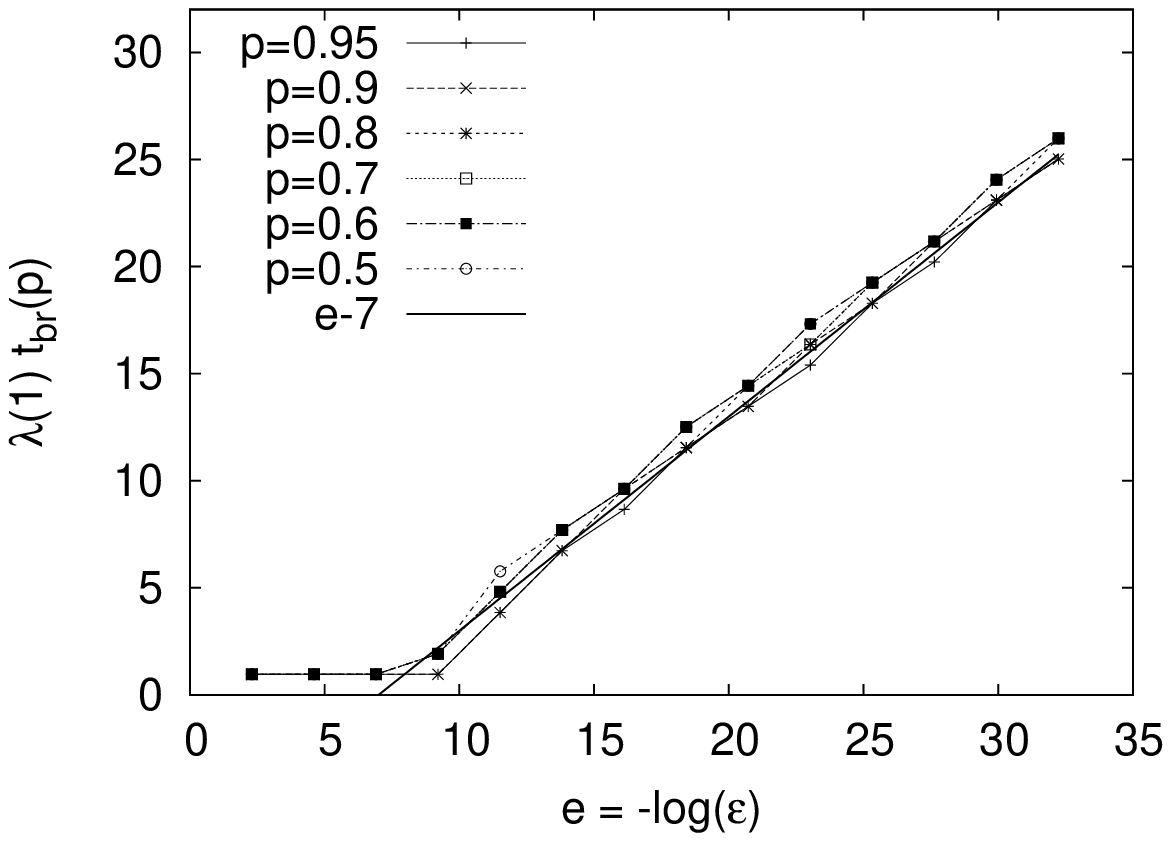}
  \hbox to15cm{\hfil(a)\hfil(b)\hfil}
\caption{The dependence of $t_{\rm br}$ on perturbation strength $\eps$ in the case of pertbation vectors $\bfm{\eps}=(0,\eps,0),~(0,0,\eps)$ (a,b) at $N=512$ and $k=1$.}
\label{pic:qcf_T}
\end{figure}

The figures \ref{pic:qcf_T_ST} and \ref{pic:qcf_T} show plots of $t_{\rm br}$ in dependence of $\eps$ for all three types of the perturbations.
Because we are discussing a discrete dynamical system, the break time $t_{\rm br}(\eps,p)$ is a discrete function of $\eps\in \bR^+$. In figure \ref{pic:qcf_T_ST} we show $t_{\rm br}$ as a function of $\eps$ in the case of non-smooth perturbation $\bfm{\eps}=(\eps,0,0)$ for two values of the classical parameter $k$.  In order to improve representation we show plots for several $p$ at the same time. We see that the heuristically obtain formula $\lambda t_{\rm br} \sim -\log \eps$ fit perfectly onto the numerical results.
The dependence of $t_{\rm br}$ on $\eps$ in the presence of smooth perturbations $\bfm{\eps}=(0,\eps,0)$ and $\bfm{\eps}=(0,0,\eps)$  is shown in figure \ref{pic:qcf_T}. We notice that the gross dependence of the break time is basically independent of perturbation.\par
\begin{figure}[!htb]
\centering
\includegraphics[width=7.5cm]{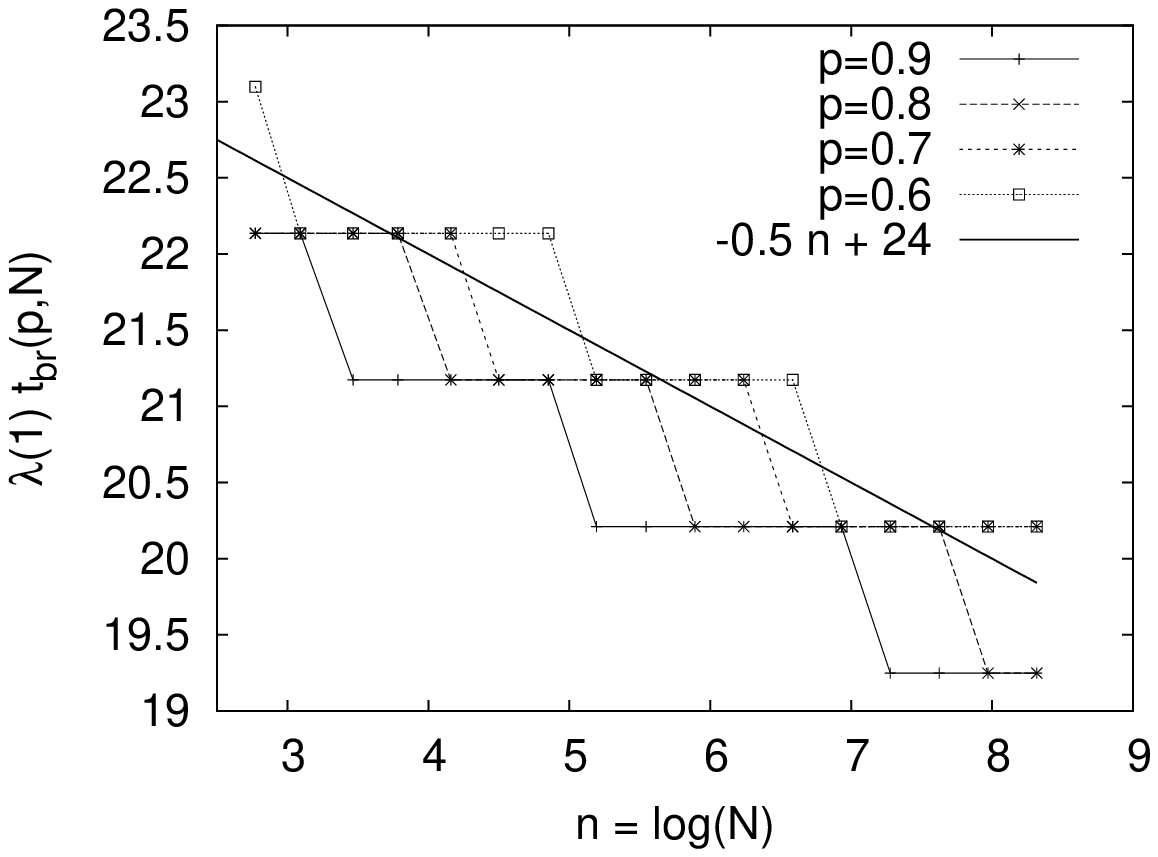}%
\includegraphics[width=7.5cm]{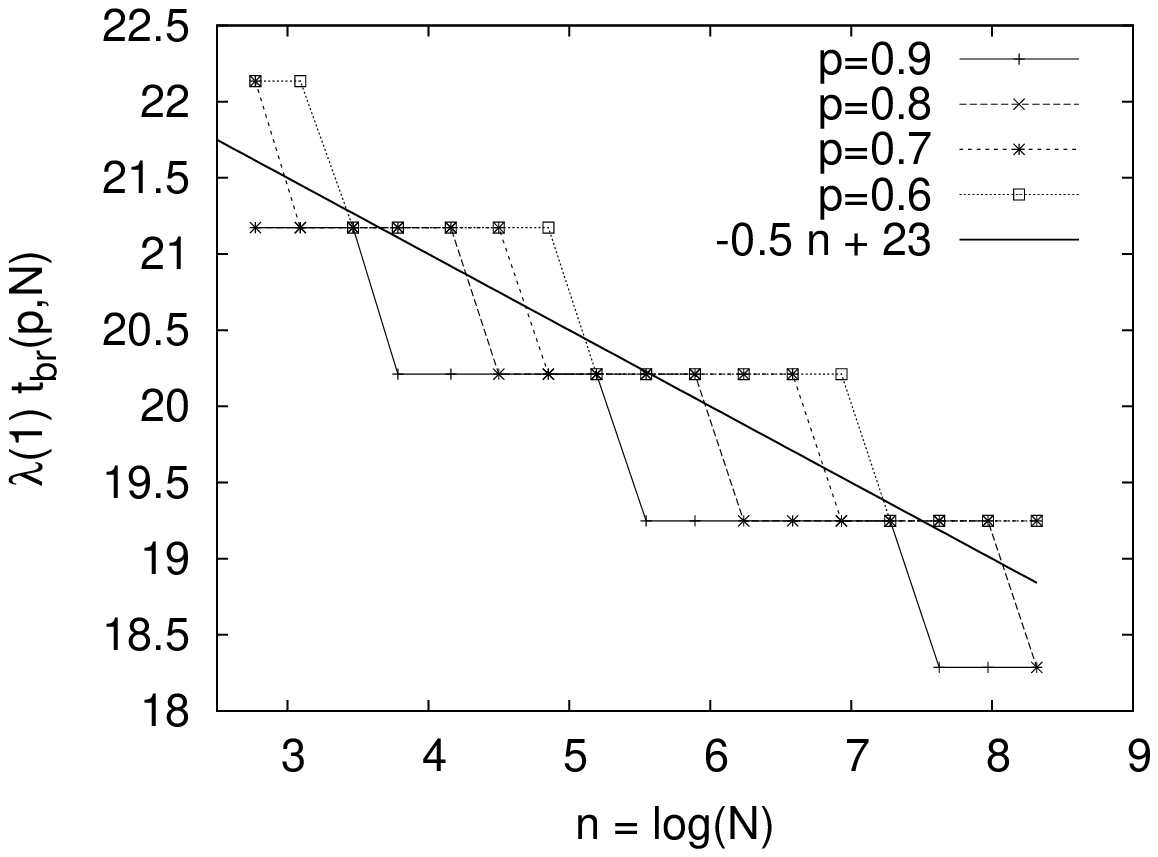}
\hbox to15cm{\hfil(a)\hfil(b)\hfil}
\caption{The dependence of $t_{\rm br}$ of Hilbert space dimension $N$ at perturbation vectors $\bfm{\eps}=(0,10^{-10},0)$, $(0,0,10^{-10})$ (a,b) for $k=1$.}
\label{pic:qcf_N}
\end{figure}

The break time $t_{\rm br}$ (\ref{eq:time_break}) depends also on the Hilbert space dimension $N$. In the limit of small perturbations $\eps N \ll 1$ we obtain from (\ref{eq:trb_gen}) the following dependence on $N$:
\beq
  \lambda t_{\rm br} \approx {\rm const.} -\frac{1}{2} \log N\>,
  \label{eq:tbr_N}
\eeq
where the constant depends on $p$, $\lambda$ and details of the initial packets.
\begin{figure}[!htb]
\centering
\includegraphics[width=7.5cm]{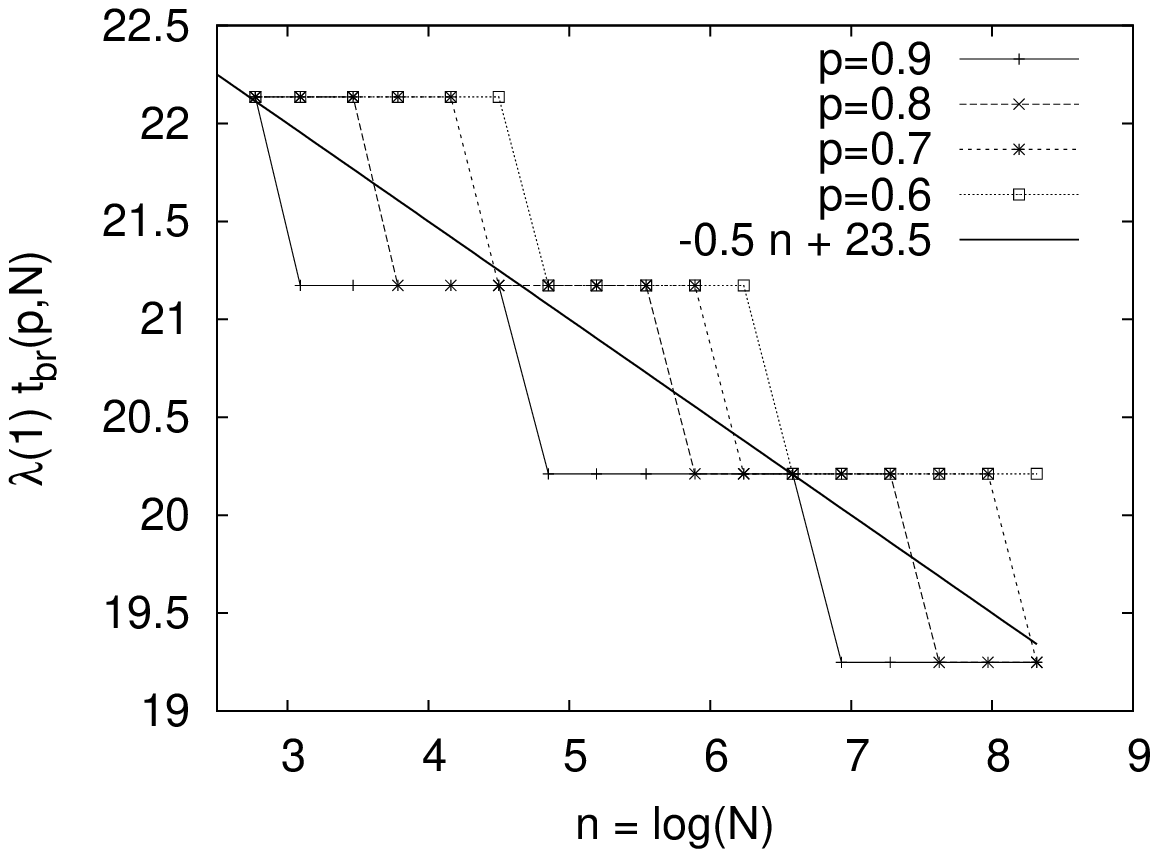}%
\includegraphics[width=7.5cm]{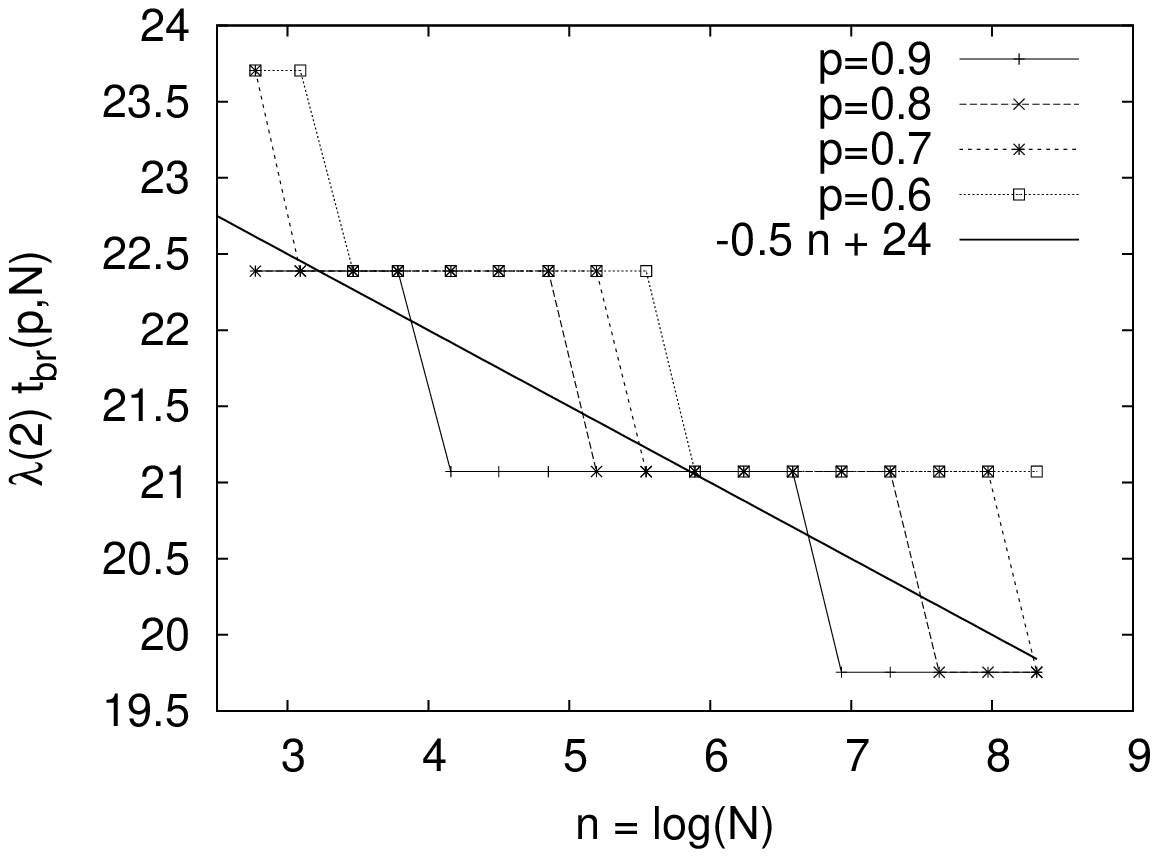}
\hbox to15cm{\hfil(a)\hfil(b)\hfil}
\caption{The dependence of $t_{\rm br}$ of Hilbert space dimension $N$ at perturbation $\bfm{\eps}=(10^{-10},0,0)$ in the case $k=1$ (a) and $k=2$ (b).}
\label{pic:qcf_N_ST}
\end{figure}
The numerical results shown in figures \ref{pic:qcf_N_ST} and \ref{pic:qcf_N} in the case of using smooth and non-smooth perturbation, respectively, confirm the theoretical dependence. But due insufficient range in variable $\log N$ we can not check the prefactor in scaling relation (\ref{eq:tbr_N}) very accurately.

Notice that the break time $t_{\rm br}$ is decreasing with increasing $N$. At the first look this would appear as a contradiction to the known QCC principle, which states that the quantum system should behave as classical system in the limit $N\to\infty$. But this is not the case: with increasing $N$ eventually $N\eps \sim 1$ and the perturbation approach becomes invalid. Thereby we enter the general regime discussed in \cite{horvat:non:qcf06}, where the break time $t_{\rm br}$ scales with $N$ as
\beq
  \lambda t_{\rm br} \asymp C \log N\>,
\eeq
where constant $C$ depends on the perturbation type. Therefore everything is still consistent with the QCC principle.

\section{Conclusions}

In this paper we investigate the correspondence between the
classical and quantum dynamics of the perturbed cat map on the torus
in the limit of semiclassical small perturbations. The
correspondence is measured by the overlap between the classical
density and the Wigner function called quantum-classical fidelity
(QCF) and denoted by $F(t)$. We study the time evolution of QCF,
which stays for a long time at the initial value $F(t)\approx 1$ and than
decays towards the ergodic value $F(t)\approx 1/N$ faster than
generally expected. The length of the initial plateau $t_{\rm br}$
scales with perturbation $\eps$ and Hilbert space dimension as
$\lambda t_{\rm br} \sim -\log(N^{\frac{1}{2}}\eps)$, where
$\lambda$ is the maximal Lyapunov exponent. At the first moment the
scaling with $N$ seem to be in  contradiction with the
correspondence principle, but this is  not the case because the
result is only meaningful for $\eps N\ll 1$. In this particular
perturbation regime, the observed behaviour is clearly far from
general and hence the  results presented here for the important and
historical model of (perturbed) cat map supplement the general
knowledge of QCC in evolving chaotic systems discussed in
\cite{horvat:non:qcf06}. It is important to notice that the
presented results can be applied to  arbitrary chaotic systems which
are {\it almost Egorov exact} or such that the Egorov exactness can
be broken by a weak perturbation.


\section*{Acknowledgments}
MH would like to thank Dipartimento di Matematica in Bologna, Italy, and Ministry of Higher education, Science and Technology of Slovenia for their financial support.
\section*{References}

\end{document}